# COEXISTENCE OF 11.2 TB/S CARRIER-GRADE CLASSICAL CHANNELS AND A DV-QKD CHANNEL OVER A 7-CORE MULTICORE FIBRE

*Emilio Hugues-Salas[1]\*, Qibing Wang[1], Rui Wang[1], Kalyani Rajkumar[1],*
*George T. Kanellos[1], Reza Nejabati[1] and Dimitra Simeonidou[1]*

[1]High Performance Networks (HPN), School of Computer Science, Electrical & Electronic Engineering and Engineering Maths. University of Bristol, Bristol, UK.
*e.huguessalas@bristol.ac.uk

## Abstract

We successfully demonstrate coexistence of record-high 11.2 Tb/s (56x200Gb/s) classical channels with a discrete-variable-QKD channel over a multicore fibre. Continuous secret key generation is confirmed together with classical channel performance below the SDFEC limit and a minimum quantum channel spacing of 17nm in the C-band.

## 1 Introduction

QKD technology is a quantum-grade encryption mechanism that has been proven to be information theoretical secure (ITS) and is the ultimate technology to strengthen the physical layer communication security. QKD has recently managed to become a commercial technology [1-2], while large deployments in Vienna, Tokyo, Battelle, Cambridge or China (Beijing-Shanghai) testbeds [3-7] have demonstrated end-to-end QKD functionality on point-to-point links. However, since current commercially deployed Discrete-Variable QKD (DV-QKD) schemes rely on the exchange of single or few photons to transfer the quantum information, they impose serious restrictions on the acceptable noise levels and the overall optical insertion losses that can be applied in a closed QKD link to survive, rendering their deployment with classical optical communication channels very difficult. Therefore, a viable co-existing scheme is needed to guarantee large scale deployment of QKD technologies in parallel to classical optical telecommunications.

To this end, several co-existing schemes that allow quantum channels and classical channels in the same optical medium have been proposed mainly relying on the principles of wavelength division multiplexing (WDM). Specifically, in [8], the use of O-Band is proposed for classical optical channels and the use of C-band for the QKD channel. This scheme is impractical for real life applications since C-Band dense wavelength-division multiplexing (DWDM) is largely deployed for the classical optical channels. Other state-of-the-art developments have demonstrated co-propagation of QKD with one 100 Gbps data channel in 150 km ultra-low loss fibre at 5 dBm launch power [9] while in [10] data transmission at Tb/s level was achieved at 11 dBm launch power. Another field trial of simultaneous QKD transmission and four 10 Gbps encrypted data channels was implemented over 26 km installed fibre at 10 dBm launch power [11]. However, as launching powers induce more noise and spectral proximity is prohibitive, WDM co-existence of classical and quantum channels is a compromise between launching powers, bandwidths and targeted secret key rates (SKR) for the quantum channels.

Another approach for achieving efficient co-existence levels lies with the use of space multiplexing techniques. In this case, multicore fibres offer enhanced channel isolation between cores and can in principle allow the unconditional co-existence of Quantum and Classical channels [12]. In this direction, multicore fibre (MCF) have been exploited so far to optimize the SKR of the quantum channel either by initiating the concept of high-dimensional quantum cryptography [13] or by exploiting the high intra-core crosstalk [14].

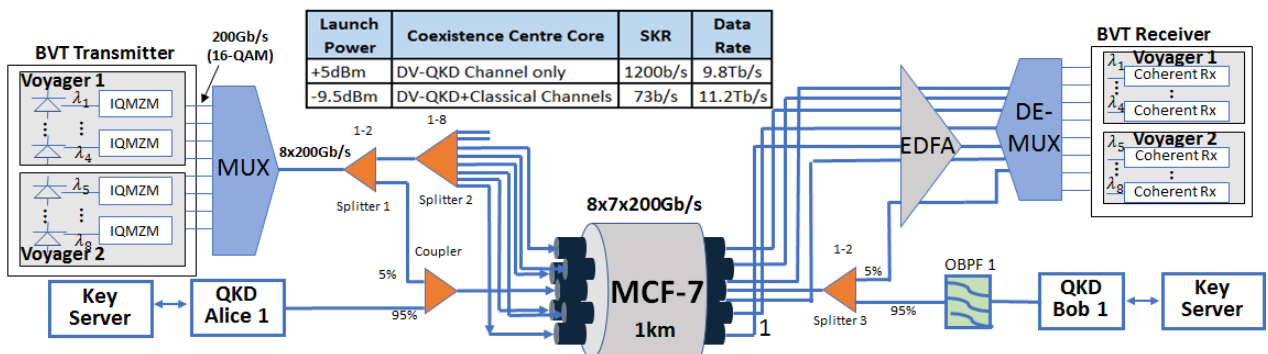

Fig. 1. Experimental Testbed for Coexistence of 11.2Tb/s of Classical Channels and a DV-QKD quantum channel over a 1km-long 7-core Multicore Fibre. Inset: Launching power levels for coexistence in MCF.

Since multi-core fibres are suitable for intra- and potentially inter- data centre deployments, short MCF links can be promoted to support intense traffic and quantum secure communications between compute nodes in data centres or between MEC clusters [15]. In a recent publication [16], we reported the preliminary results to optimize the transmission capacity of the classical optical channels in the peripheral cores of an MCF while sustaining the viability of a single quantum channel in the central core.

In this paper, we generalize our study by allowing a quantum channel and multiple classical channels in all the cores of the MCF, leading to a new coexistence transmission record of 11.2Tb/s for the classical channels and a maximum secret key rate (SKR) of 920 b/s and a QBER of 3.7% for the DV-QKD channel coexisting with the classical channels in the same core with a minimum quantum channel spacing of 17nm. Moreover, during the experimental demonstration it was found a carrier-grade optical launching power of +5dBm for the classical channels over the MCF cores without coexisting quantum and classical channels in the centre core, resulting in a minimum QBER of 2.5% and a SKR of 1200b/s.

## 2  Experimental System Setup

Fig. 1 shows the experimental testbed for the coexistence of 11.2 Tb/s of classical channels and a DV-QKD quantum channel over a MCF. Classical channels are generated by two Facebook Voyager switches with bandwidth variable transponders (BVTs). Each switch consists of 4 BVTs with the capability to configure modulation format (DPSK, 8-QAM and 16-QAM) and wavelength within the C-band region. For this demonstration, each wavelength of the voyager is configured with 16-QAM modulation format for a maximum data rate of 200 Gb/s per channel. These generated classical channels are connected to a wavelength selective switch (WSS) that multiplexes the 8 channels from both Voyagers, increasing the throughput to 1.6 Tb/s and feeding a 3-dB splitter to divide the multiplexed signal into two paths. One path is connected to a 1/8 splitter to equally distribute classical channels over 6 cores of the MCF. The second path is coupled into the quantum channel via a 5/95% coupler. Therefore, every core in the MCF has a data rate of 1.6 Tb/s, leading to a total transmitted data rate of 11.2 Tb/s over the MCF.

After transmission over the 1-km MCF, with average loss of 2.7dB per core, the DWDM signal is amplified, demultiplexed by a WSS, and transferred to the coherent receivers of the Voyager switches where BER and received optical power are measured. The quantum channel is split for exchange of encoded photons and BER measurement by using another 5/95% splitter. This quantum channel has a total optical power budget of 6.7dB end-to-end (Alice to Bob). Before decoding the classical signals after the coherent receiver included in the Voyager switch, their optical power is pre-amplified by an EDFA to compensate for the power losses of the MCF and WSS. With respect to the quantum channel, a 2nm-bandwidth optical filter is used just before the QKD-Bob unit to filter out any leakage from the classical signals, maintaining the optical total link power budget for the quantum channel below 10dB specified for the QKD system.

## 3  DV-QKD Channel and 11.2 Tb/s Classical Coexistence over a 7-Core Multicore Fibre

*3.1 Spectrum of the Quantum and Classical Channel Coexistence over a Single Core of the Multicore Fibre*

Fig. 2 shows the spectrum of the 8x200Gb/s classical channels coexisting with the quantum channel over the central core of the MCF. As illustrated in Fig. 2, the channel spacing in between the quantum and the classical channels is 17nm, observed by using a tap coupler in the quantum channel. The optical filter profile used to filter out the Raman noise generated by the classical channels to enable the photon-exchange is also shown in Fig. 2. In addition, the spectra of the classical channels from an adjacent core of the MCF are also illustrated in Fig. 2 before demultiplexing in the receiver.

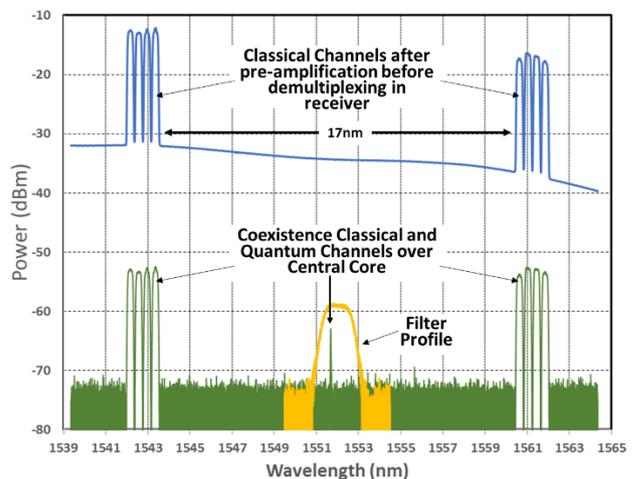

Fig. 2. Spectrum of the combined transmission over a single core of quantum and classical channels.

*3.2 Channel Spacing between Classical and Quantum Channel for Coexistence over a Single Core*

To further investigate the effect of the classical channels over the quantum channel over the same core, Fig. 3 illustrates the performance of the coexisting channels by decreasing the spectral spacing between them. Fig. 3 shows a minimum spacing of 17nm between non-contiguous channels with a QBER of 5.7% and a secret key rate of 100b/s (quantum channel), and an average BER of $1.8 \times 10E^{-3}$ (classical channel) being significantly lower than the Soft-Decision FEC limit enabled by the Voyager switch. Beyond this minimum spacing, the QKD system stops generating keys. In here, the transmission over the other cores was performed in parallel without the performance of these channels being affected, since the crosstalk of the MCF enabled the coexistence for 11.2Tb/s. Also, the direction of transmission of the classical channels is counter-propagating the transmission of encoded photons from QKD-Alice to QKD-Bob units.

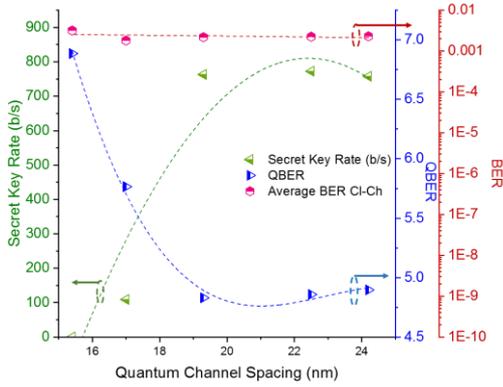

Fig.3. SKR, QBER and BER vs channel spacing between the quantum and classical channels of the MCF central core.

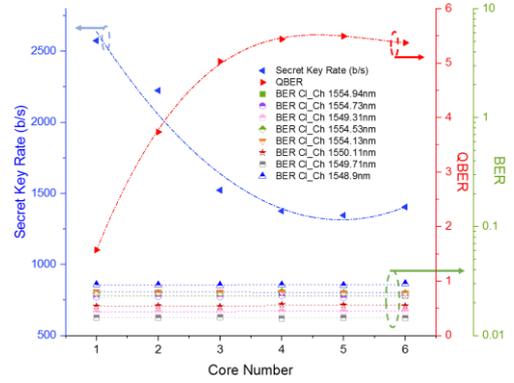

Fig. 5. SKR, QBER and BER vs combined number of cores with classical channels.

*3.3 Performance Evaluation of the Coexistence of a DV-QKD Channel and 11.2Tb/s Classical Channels over a 7-core Multicore Fibre*

Fig. 4 shows the curve of the SKR, QBER and BER vs launch power for the transmission of eight 16-QAM-modulated optical channels and a DV-QKD channel over the central core of the MCF. To increase the launched power, the power levels of individual channels were increased from the BVT interfaces. As observed in Fig. 4, the QBER increases and the SKR decreases when the launched optical power into the MCF is increased up to a maximum power of -9.5dBm. Beyond this limit, the system stops generating keys due to excessive Raman noise leakage over the quantum channel. On the classical channels, a maximum average BER of $6.68 \times 10^{-4}$ was obtained for the maximum launched power of -9.5dBm.

## 4 Conclusion

The coexistence of a DV-QKD channel and 56x200Gb/s classical channels was successfully demonstrated over a 1-km-long multicore fibre for a record-high coexistence transmission of 11.2Tb/s. For the coexistence over the central core of the MCF, a minimum QBER of 3.7% and a maximum SKR of 920b/s was demonstrated for the DV-QKD simultaneously with a minimum average BER of $1.28 \times 10^{-2}$ for the classical channels. In addition, investigations prove that a minimum channel spacing of 17nm is required in between quantum and classical channels and that the incremental addition of classical signals at different cores will degrade the quantum channel for a maximum QBER (SKR) of 5.3% (1400b/s). This work shows that DV-QKD can effectively coexist with carrier-grade equipment over MCF for maximum capacity and continuous key generation, being suitable for secure intra- and inter-data centre applications as well as MEC.

## 5 Acknowledgements

This work acknowledges the EU Horizon 2020 UNIQORN, the project EPSRC EP/M013472/1: UK Quantum Hub for Quantum Communications Technologies and the EP/L020009/1: Towards Ultimate Convergence of All Networks. The authors also thank the support from Facebook for the Voyager switches in the experiment.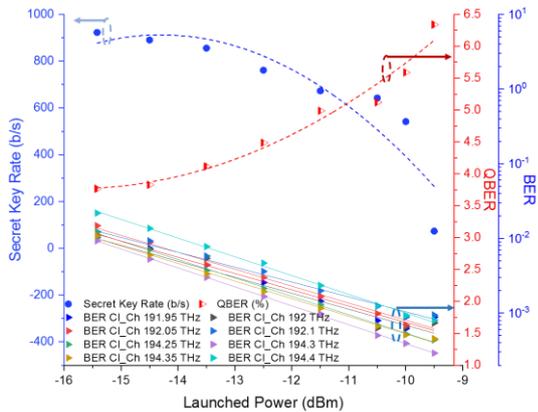

Fig. 4. SKR, QBER and BER (Central Core) vs launched optical power into a 7-core multicore fibre.

To explore the impact of the coexistence from adjacent cores of the MCF, Fig. 5. shows the performance of the quantum and classical channels after adding channels incrementally at different cores. As observed, for the first two added cores, the quantum channel QBER and SKR are kept within the limits of 3.5% and 2220b/s, respectively. Beyond three added cores, the performance degrades to 5% (QBER) and 1520b/s (SKR). However, the same performance is kept when adding classical signals for the total six additional cores.